# Sensitivity estimation for calculated phase equilibria

Richard Otis[1,*], Brandon Bocklund[2], Zi-Kui Liu[2]

[1] Engineering and Science Directorate, Jet Propulsion Laboratory, California Institute of Technology, Pasadena, CA 91109, USA

[2] Department of Materials Science & Engineering, Pennsylvania State University, University Park, PA, 16802, USA

**Abstract**

The development of a consistent framework for Calphad model sensitivity is necessary for the rational reduction of uncertainty via new models and experiments. In the present work, a sensitivity theory for Calphad was developed, and a closed-form expression for the log-likelihood gradient and Hessian of a multi-phase equilibrium measurement was presented. The inherent locality of the defined sensitivity metric was mitigated through the use of Monte Carlo averaging. A case study of the Cr-Ni system was used to demonstrate visualizations and analyses enabled by the developed theory. Criteria based on the classical Cramér–Rao bound were shown to be a useful diagnostic in assessing the accuracy of parameter covariance estimates from Markov Chain Monte Carlo. The developed sensitivity framework was applied to estimate the statistical value of phase equilibria measurements in comparison with thermochemical measurements, with implications for Calphad model uncertainty reduction.

* Corresponding author. E-mail: richard.otis@jpl.nasa.gov

**Introduction**

Calphad-based thermodynamic models are routinely used to probe the phase stability in multi-component systems. Computational efficiency and the ability to incorporate experimental measurements, atomistic simulations, and expert intuition in a semi-empirical fashion have led to the broad adoption of the Calphad approach, but it is only in recent years that serious attention has been paid to uncertainty quantification (UQ) of the model predictions.

Stan and Reardon demonstrated early work on Calphad UQ using genetic algorithms which anticipated the Bayesian approach adopted by later work [1]. As computing efficiency increased, several authors identified Markov Chain Monte Carlo (MCMC) as a powerful technique for optimizing Calphad model parameters and simultaneously determining their uncertainty with respect to the data [2,3]. Readers interested in further discussion of recent developments in Bayesian UQ for ICME, with application to Calphad, are directed to a recent review [4].

While there have been significant developments in understanding the propagation of Calphad model uncertainty to the equilibrium predictions [5], the inverse has received minimal attention since the work of Jansson in Calphad parameter optimization [6]. In seeking to develop a theory of Calphad model sensitivity, it is desired to understand the flow of uncertainty from the experimental measurements to a given Calphad model. Clear definitions must be given to all observation types, including multi-phase equilibrium information, commonly referred to as "phase diagram data." The development of a consistent framework for Calphad model sensitivity is necessary, not only for UQ, but for the rational reduction of uncertainty via new models and experiments.

**Theory of Calphad Model Sensitivity**

Without loss of generality to multi-component, multi-sublattice systems, we will first consider an isobaric binary system, A-B, consisting of two single-sublattice phases, α and β. The molar Gibbs energies are defined as $G_m^\alpha(T, y_A^\alpha, y_B^\alpha; \boldsymbol{\theta})$ and $G_m^\beta(T, y_A^\beta, y_B^\beta; \boldsymbol{\theta})$, respectively, where $T$ is the temperature, $y_i^j$ is the site fraction of component $i$ in phase $j$, and $\boldsymbol{\theta}$ is an empirically-determined vector of continuously-valued model parameters for the phases. $G_m^j$ may have a non-linear dependence on the elements of $\boldsymbol{\theta}$, but the partial derivatives with respect to the parameters are assumed to exist. It is often the case that each element of $\boldsymbol{\theta}$ is associated with only one phase, but that assumption is not necessary. $N_j$ is the molar amount of phase $j$. The total molar amount of components A and B, $M_A$ and $M_B$, are equal to $N_\alpha y_A^\alpha + N_\beta y_A^\beta$ and $N_\alpha y_B^\alpha + N_\beta y_B^\beta$, respectively.

In the interest of brevity, detailed solutions for the equations of equilibrium are not included here. This derivation can be found in several publications, recently in significant detail by Sundman et al. [7]. For this work it was sufficient to assume a particular solution is known to the equations under the given conditions.

Let an experimental observation at equilibrium of the coexistence of phases $\alpha$ and $\beta$ at a fixed temperature $T$ find the measured mole fractions $\tilde{x}_A^\alpha$, $\tilde{x}_B^\alpha$, $\tilde{x}_A^\beta$ and $\tilde{x}_B^\beta$. The mole fractions predicted by the candidate model specified by $\boldsymbol{\theta}$ are $x_A^\alpha$, $x_B^\alpha$, $x_A^\beta$ and $x_B^\beta$. The "residual driving force" formulation for the error in a candidate model specified by $\boldsymbol{\theta}$ of a multi-phase equilibrium observation was adopted in this work [8,9]. The Gibbs energies of phases in coexistence lie on an equipotential line (hyperplane in multi-component systems) which minimizes the total energy. When the model degrees of freedom, $\boldsymbol{\theta}$, do not satisfy that criteria for a given experimental measurement, the deviation from the equipotential condition can be quantified as a signed distance ("driving force") from a fictitious hyperplane. That hyperplane is the arithmetic mean of the hyperplanes calculated at the measured phase compositions using the candidate model. As the deviation approaches zero, the hyperplanes at each measured composition will approach the mean (Figure 1).

An advantage of this formulation is that it is continuous with respect to the metastability of one or more observed phases, meaning that if an observed phase is not predicted to be stable according to the candidate model, the error can still be defined. Another advantage is that it is differentiable with respect to $\boldsymbol{\theta}$. The "residual driving force" of a multi-phase equilibrium measurement is defined as

$$R(\boldsymbol{\theta}) = \sum_i \left(\bar{\mu}_i \tilde{x}_i^j\right) - \sum_i \left(\mu_i^j \tilde{x}_i^j\right) \qquad (1)$$

The index $j$ refers to each observed phase, i.e., $\alpha$ or $\beta$. Equation 1 is computed at $\tilde{x}_i^j$, the tie-line endpoint corresponding to each observed phase $j$. $\bar{\mu}_i$ is the arithmetic mean of the chemical potentials found by computing a *multi-phase* global equilibrium at each tie-line endpoint, i.e., an average chemical potential of values calculated from their respective compositions at all tie-line endpoints. For the $\mu_i^j$ terms, corresponding to the individual phases, we compute *single-phase* local equilibria, i.e., compute the chemical potentials at $\tilde{x}_i^\alpha$ while only considering the $\alpha$ phase, and also the chemical potentials at $\tilde{x}_i^\beta$ while only considering the $\beta$ phase (Figure 1). When $\boldsymbol{\theta}$ are chosen such that $\tilde{x}_i^j = x_i^j$, $R(\boldsymbol{\theta})$ will be equal to zero. It is often the case that, in an experimental multi-phase equilibrium observation, only one of the phase compositions can be determined, e.g., a measurement of the liquidus temperature by differential scanning calorimetry heating/cooling curve analysis. For the case of a phase $\beta$ of undetermined composition in equilibrium with a phase $\alpha$ at measured composition $\tilde{x}_i^\alpha$, $\tilde{x}_i^\beta$ can be estimated as the composition which maximizes the thermodynamic driving force for formation of the $\beta$ phase relative to $\bar{\mu}_i$. The computation of $\bar{\mu}_i$ then excludes the chemical potentials calculated at the estimated $\tilde{x}_i^\beta$.

For sensitivity estimation, the first derivatives of $R(\boldsymbol{\theta})$ need to be computed. Assume that $x_i^j$ is independent of $\boldsymbol{\theta}$. The derivative can then be written as follows.

$$\frac{\partial R}{\partial \boldsymbol{\theta}} = \sum_i \left( \frac{\partial \bar{\mu}_i}{\partial \boldsymbol{\theta}} \tilde{x}_i^j \right) - \sum_i \left( \frac{\partial \mu_i^j}{\partial \boldsymbol{\theta}} \tilde{x}_i^j \right) \tag{2}$$

Because the arithmetic mean is a linear operator, it is sufficient to determine an expression for $\frac{\partial \mu_i}{\partial \boldsymbol{\theta}}$. The chemical potentials are dependent variables which are outcomes of a non-linear optimization. The Lagrangian formulation of the Gibbs energy minimization problem [6] [7] can be stated as follows.

$$H_L = \sum_j N^j G_m^j - \sum_i f_i \mu_i - \sum_l c_l \lambda_l \tag{3}$$

$f_i = M_i - \widetilde{M}_i$ is the mass balance constraint for component $i$, $l$ is an index for the internal constraints $c_l$ (e.g., site fraction balance) for all phases in equilibrium, where $\lambda_l$ is the corresponding Lagrange multiplier. For the following and all subsequent steps, we assume calculation at a feasible solution, so that the gradient of the Lagrangian is equal to zero. For the present case this can be expanded as the following system of equations:

$$\begin{bmatrix} \frac{\partial c_l(y_k)}{\partial y_k} & \frac{\partial f_A}{\partial y_k} & \frac{\partial f_B}{\partial y_k} \\ 0 & \frac{\partial f_A}{\partial N^j} & \frac{\partial f_B}{\partial N^j} \end{bmatrix} \begin{bmatrix} \lambda_l \\ \mu_A \\ \mu_B \end{bmatrix} = \begin{bmatrix} \frac{\partial (\sum_j N^j G_m^j)}{\partial y_k} \\ G_m^j \end{bmatrix} \quad (4)$$

$y_k$ are the internal variables for all stable phases in the calculation, and $N^j$ is the amount of phase $j$. $\lambda_l$ can otherwise be discarded for the present analysis.

Assume that all $c_l$ and $M_i$ are independent of $\boldsymbol{\theta}$ and differentiate the system of equations, resulting in

$$\begin{bmatrix} \frac{\partial c_l(y_k)}{\partial y_k} & \frac{\partial f_A}{\partial y_k} & \frac{\partial f_B}{\partial y_k} \\ 0 & \frac{\partial f_A}{\partial N^j} & \frac{\partial f_B}{\partial N^j} \end{bmatrix} \begin{bmatrix} \frac{\partial \lambda_l}{\partial \boldsymbol{\theta}} \\ \frac{\partial \mu_A}{\partial \boldsymbol{\theta}} \\ \frac{\partial \mu_B}{\partial \boldsymbol{\theta}} \end{bmatrix} = \begin{bmatrix} \frac{\partial (\sum_j N^j G_m^j)}{\partial y_k \partial \boldsymbol{\theta}} \\ \frac{\partial G_m^j}{\partial \boldsymbol{\theta}} \end{bmatrix} \quad (5)$$

Because of redundant constraints, under some circumstances this system of equations may be over-determined. By adopting the least-squares solution and referring to the preceding matrix as $A$, the chemical potential $\boldsymbol{\theta}$ derivatives can be written in closed form:

$$\begin{bmatrix} \frac{\partial \lambda_l}{\partial \boldsymbol{\theta}} \\ \frac{\partial \mu_A}{\partial \boldsymbol{\theta}} \\ \frac{\partial \mu_B}{\partial \boldsymbol{\theta}} \end{bmatrix} = (A^T A)^{-1} A^T \begin{bmatrix} \frac{\partial (\sum_j N^j G_m^j)}{\partial y_k \partial \boldsymbol{\theta}} \\ \frac{\partial G_m^j}{\partial \boldsymbol{\theta}} \end{bmatrix} \quad (6)$$

This expression shares some similarity with the temperature "dot derivative" in Sundman et al. [7], if the model degrees of freedom ($\boldsymbol{\theta}$) are mathematically interpreted as independent thermodynamic state variables, similar to temperature. It is often the case that the Gibbs energy model of a phase has a linear dependence on $\boldsymbol{\theta}$, e.g., $G_m^j = \cdots + y_A y_B (\theta_1 + \theta_2 T)$. For this case, $\frac{\partial G_m^j}{\partial \boldsymbol{\theta}}$ is independent of $\boldsymbol{\theta}$ and is constant for a given equilibrium solution, assuming the phases' internal degrees of freedom do not change. $\frac{\partial \mu_i}{\partial \boldsymbol{\theta}}$ is also constant, as a consequence. If this scenario applies to all the phases in the calculation, then $\frac{\partial R}{\partial \boldsymbol{\theta}}$ will be independent of $\boldsymbol{\theta}$, and $R(\boldsymbol{\theta})$ will be linear in $\boldsymbol{\theta}$.

**Definition of Phase Equilibria Log-Likelihood**

Assume that the error associated with an experimental observation of $R(\boldsymbol{\theta};\ \tilde{x}_i^j)$ is normally distributed about zero with constant variance $\sigma^2$. Note that $R(\boldsymbol{\theta};\ \tilde{x}_i^j)$ is not a true observable because its value is inferred from the measured quantities $\tilde{x}_i^j$, but in principle this only affects computation of $\sigma^2$. The log-likelihood function for $p$ independent observations can then be expressed as:

$$\ln L(\boldsymbol{\theta}) = -\sum_p \frac{1}{2\sigma_p^2} \| R_p(\boldsymbol{\theta};\ \tilde{x}_i^{j,p}) \|^2 - \sum_p \frac{1}{2} \ln\left(\frac{2\pi}{\sigma_p^2}\right) \qquad (7)$$

The second term is often omitted, as it is independent of $\boldsymbol{\theta}$. In the present work attention is restricted solely to multi-phase equilibrium observations, though observations of other thermochemical quantities, such as heat capacity and enthalpy of formation, can easily be incorporated in the log-likelihood and the following derivation of sensitivity. In statistics, the score function $s(\boldsymbol{\theta})$ is defined as the gradient of the log-likelihood function.

$$s(\boldsymbol{\theta}) \stackrel{\text{def}}{=} \frac{\partial \ln L(\boldsymbol{\theta})}{\partial \boldsymbol{\theta}} = -\sum_p \frac{R_p(\boldsymbol{\theta};\ \tilde{x}_i^{\alpha,p}, \tilde{x}_i^{\beta,p})}{\sigma_p^2} \frac{\partial R_p(\boldsymbol{\theta};\ \tilde{x}_i^{\alpha,p}, \tilde{x}_i^{\beta,p})}{\partial \boldsymbol{\theta}} \qquad (8)$$

A "partial" score of a particular observation can also be defined and is denoted $s_p(\boldsymbol{\theta})$. By the independence of observations, $s(\boldsymbol{\theta}) = \sum_p s_p(\boldsymbol{\theta})$.

The corresponding Fisher information matrix (FIM), $I(\tilde{x}_i^{j,p}; \boldsymbol{\theta})$, captures the curvature of the log-likelihood, and is defined as the negative of the expectation of the log-likelihood Hessian as follows

$$I(\tilde{x}_i^{j,p}; \boldsymbol{\theta}) \stackrel{\text{def}}{=} -E\left(\frac{\partial^2 \ln L(\boldsymbol{\theta})}{\partial \boldsymbol{\theta}^T \partial \boldsymbol{\theta}}\right) = \sum_p \frac{1}{\sigma_p^2}\left[\frac{\partial^2 R_p(\boldsymbol{\theta})}{\partial \boldsymbol{\theta}^T \partial \boldsymbol{\theta}} R_p(\boldsymbol{\theta}) + \frac{\partial R_p(\boldsymbol{\theta})}{\partial \boldsymbol{\theta}} \frac{\partial R_p(\boldsymbol{\theta})}{\partial \boldsymbol{\theta}^T}\right] \quad (9)$$

Under a common condition, discussed in the previous section, that $R_p(\boldsymbol{\theta})$ is linear in $\boldsymbol{\theta}$, $\frac{\partial^2 R_p(\boldsymbol{\theta})}{\partial \boldsymbol{\theta}^T \partial \boldsymbol{\theta}} = 0$ and the higher-order term can be neglected, i.e.,

$$I(\tilde{x}_i^{j,p}) = \sum_p \frac{1}{\sigma_p^2} \frac{\partial R_p(\tilde{x}_i^{j,p})}{\partial \boldsymbol{\theta}} \frac{\partial R_p(\tilde{x}_i^{j,p})}{\partial \boldsymbol{\theta}^T} \quad (10)$$

The assumption of linearity causes the dependence on $\boldsymbol{\theta}$ to fall out of Equation 10 because $\frac{\partial R_p(\boldsymbol{\theta})}{\partial \boldsymbol{\theta}}$ is constant in $\boldsymbol{\theta}$, and so the FIM is purely a function of the underlying model form and the experimentally observed $\tilde{x}_i^{j,p}$. There is an important caveat: In this toy problem there are only two phases, but in multi-component systems with several phases, a phase equilibria measurement could involve only a subset of the potentially-stable phases in a given system. If an experimental measurement only observes, e.g., the phases α and β, and a candidate thermodynamic model also only predicts α and/or β phase(s) under the same conditions, then small perturbations of the parameters of a third phase, γ, have zero effect on the log-likelihood of that observation. More generally, for the case where none of the phases which are observed or predicted have a dependence on a particular element of $\boldsymbol{\theta}$ ($\theta_m$), $\frac{\partial R_p(\boldsymbol{\theta})}{\partial \theta_m} = 0$. This result is intuitive, since an observation cannot provide any information about the value of a parameter, when the model does not depend on that parameter. However, if a candidate model mis-predicts the presence of a phase γ, an experimental observation of α/β equilibrium under the same conditions defines a log-likelihood that is locally a function of the parameters of all three phases.

A scalar sensitivity metric based on the FIM can be defined several ways, and is strongly connected to the notion of the optimality of a measurement. In the present work a form of "A-optimality" was adopted, wherein the trace of the FIM was used to define the sensitivity [10].

$$S(\tilde{x}_i^{j,p}) \stackrel{\text{def}}{=} \operatorname{tr} I(\tilde{x}_i^{j,p}) = \sum_m \sum_p \frac{1}{\sigma_p^2} \left\| \frac{\partial R_p(\tilde{x}_i^{j,p})}{\partial \theta_m} \right\|^2 \tag{11}$$

In Calphad modeling the numerical values of the model parameters can vary over several orders of magnitude (depending on whether the parameter is multiplying a constant, $T^3$, higher-order interaction, etc.), complicating sensitivity comparisons involving different parameters. The "scaled sensitivity" $Z(\tilde{x}_i^{j,p})$ can be understood as a measure of how much the specified observations help reduce the variance of $\boldsymbol{\theta}$. This dimensionless scalar quantity has the desirable property of being additive in observations as well as in parameters, facilitating sensitivity comparisons between subgroups of observations and/or parameters.

$$Z(\tilde{x}_i^{j,p}) = \sum_m \sum_p \frac{\sigma_m^2}{\sigma_p^2} \left\| \frac{\partial R_p(\tilde{x}_i^{j,p})}{\partial \theta_m} \right\|^2 \tag{12}$$

The specific definition in Equation 12 of an observation, $p$, warrants discussion. A "phase region" is defined in the present work as a group of tie-line endpoints corresponding to the same multi-phase equilibrium. The approach taken in this work was to define an observation in terms of each measured phase region, such that each dataset consisted of multiple "observations," all assumed statistically independent. This definition preserved the additivity property of $Z(\tilde{x}_i^{j,p})$ and made it convenient to generate sensitivity analyses based on both the parameters and the datasets ("groups of phase regions"). A disadvantage of this approach was that trends in $Z(\tilde{x}_i^{j,p})$ with respect to the MCMC iterations were challenging to interpret. The MCMC simulation involves a maximization of the log-likelihood function (Eq. 7) and, while it is expected that the magnitude of the total log-likelihood gradient (Eq. 8) decreases with an approach toward the maximum-likelihood value of $\boldsymbol{\theta}$, the same is not generally true for $Z(\tilde{x}_i^{j,p})$. This is because of error cancellation due to a summation of terms with opposing sign in the log-likelihood gradient. In the scaled sensitivity as defined in the present work, the gradient of *each observation* is squared prior to summation, so opposing gradients do not cancel. For the case of strongly-conflicting (inconsistent) observations, the value of $Z(\tilde{x}_i^{j,p})$ may be large, even near the maximum-likelihood $\boldsymbol{\theta}$. Another possibility would have been to define each dataset as a single "observation," such that gradients cancel in a way similar to Equation 8. That approach could have value as a diagnostic quantity for Calphad modeling during the parameter optimization process, but a scaled sensitivity defined in such a way would be strongly correlated with the log-likelihood gradient, and so such an analysis might be better performed by just computing the gradient norm. While the issue of potentially conflicting data has been investigated in the context of the pure elements [11], further analysis of the role of outliers in multi-component Calphad sensitivity analyses is recommended for future work.

A potential limitation of the present approach is that the sensitivity metric is only local and, for example, a "nearly mis-predicted" phase close to the limit of stability for a given tie-line will contribute nothing to the sensitivity, even though a small change in $\boldsymbol{\theta}$ could cause it to become stable in the measured phase region of interest. This concern was mitigated through the use of Monte Carlo estimates of $\boldsymbol{\theta}$ around the maximum-likelihood value. $Z(\boldsymbol{\theta}; \tilde{x}_i^{j,p})$ was then computed as chain/trace averages, which introduced a degree of non-locality to the predictions.

Another potential limitation is that correlations between parameters are not explicitly considered, i.e., the covariance of $\boldsymbol{\theta}$. While it was not done in the present work, other optimality criteria incorporating the off-diagonal elements of Equation 10 are known [10]. One challenge to resolve for such an approach would be finding good empirical estimates of the covariance to use in the rescaling of the FIM.

**Application to Cr-Ni**

The Cr-Ni system is a common exemplar system for thermodynamic modeling of alloys, given its technological importance and relative simplicity. It contains three stable solution phases: fcc, bcc and liquid. If one adopts the Standard Element Reference and tabulated lattice stabilities for the solution phases [12], then it is only left to the modeler to determine the binary interaction parameters for each phase. The low-temperature intermetallic phase, $CrNi_2$, is neglected in the present work, as are all magnetic contributions. A complete thermodynamic assessment was not an objective of this work, and such interested readers are directed to a recent review by Tang and Hallstedt [13].

The initial Cr-Ni thermodynamic model was generated by the ESPEI software using thermochemical data (enthalpies, entropies) for the individual phases [8]. The iterative least-squares procedure used by ESPEI generated a database with non-zero $a + bT$ terms for the Redlich-Kister binary interactions of both degree 0 and 1, in all three considered phases, for a total of 12 adjustable model parameters. The naming convention for the binary interaction parameters was L(*phase*;*Redlich-Kister degree*)[A,B], depending on which coefficient of the corresponding $a + bT$ expression was being referenced. The phase diagram of this initial model is shown in Figure 2(a). The thermochemical data used to generate the starting point was then discarded for the subsequent analysis. Typically this data would be retained in a Calphad modeling procedure, but it was excluded in this work to isolate the statistical influence of the phase equilibria measurements.

For the MCMC step, phase equilibria data for the solution phases was collected from the literature [14–25]. The ESPEI YAML configuration file and JSON data files for the MCMC step can be found in the Supplementary Material. 24 chains (twice the number of parameters) were included in the ensemble. All data were equally weighted with an ESPEI "data weight" of 20, for an effective $\sigma_p$ of 50 J/mol. A flat prior, contributing a log-prior of zero to the log-probability, was assumed for all parameter values. The MCMC simulation was run for 500 fixed iterations without stopping criteria. The phase diagram from the chain-average parameters at the last iteration is shown in Figure 2(b). The log-likelihood trace is shown in Figure 3.

Details on the code and data files needed to reproduce the figures and the table can be found in the Supplementary Material.

An after-the-fact sensitivity analysis was conducted on a Cr-Ni thermodynamic model using the parameter trace from the MCMC simulation. Log-likelihood gradients and scaled sensitivities were computed according to Equations 8 and 12, respectively, and stored for each experimental measurement at each MCMC iteration. The scaled sensitivity was computed as a summation over the observations ($p$) contained within each dataset, and then normalized based on the number of contained measurements (phase regions). The empirical variance of each parameter, $\sigma_m^2$, was computed based on the trace of the last 300 iterations, marginalized over all chains.

Figure 4 shows the dataset scaled sensitivity per phase region. The scaled sensitivity (Eq. 12) was computed for each dataset and then normalized based on the number of contained measurements (phase regions). For most datasets, there was a general decreasing trend with the number of Markov Chain Monte Carlo (MCMC) iterations, until settling around a particular value. A significant increase in scaled sensitivity was seen from the Bechtoldt1961 dataset (defining A1 phase compositions at the Ni-rich boundary with A2). This was understood to be caused by within-dataset disagreement on the sign of the log-likelihood gradient and could be an indicator of inconsistent observations, insufficient degrees of freedom in the present model, or both.

The computed sensitivities can also be visualized in terms of each model degree of freedom. The contribution of each parameter to the scaled sensitivity is shown as a function of MCMC iterations in Figure 5. The sensitivity contribution from the higher-order liquid parameters was minimal throughout the optimization process, indicating that the considered observations were relatively uninformative for those model degrees of freedom. High sensitivities seen from the higher-order parameters in the A1 phase were consistent with the dataset-based analysis (Figure 4), and were understood as an indicator that those parameters were strongly coupled to the observations, particularly at the Ni-rich side of the A1-A2 phase boundary.

Sensitivities can also be projected back into the space of the observations, providing insight into where new measurements might make the most impact on the likelihood. Figure 6 shows the scaled sensitivity per parameter averaged over the last 300 MCMC iterations, visualized in the space of observations. The result comported with intuition, with parameters showing sensitivity to the phase equilibria measurements from the corresponding phase. Some parameters showed sensitivity corresponding to the "far" ends of equilibrium tie-lines involving the given phase, consistent with the coupling between phases defined by Equation 1. In attempting to explain the compositional sensitivity fluctuation observed in the higher-order liquid parameters, consider that binary Redlich-Kister parameters of degree 1 achieve extreme values at mole fractions of approximately 0.21 and 0.79. This could explain the localized sensitivity peak observed in the higher-order liquid parameters, though the absolute magnitude of the sensitivity for those parameters was still observed to be small.

Highly-focused analyses became possible with data at this resolution, enabling consideration of the impact of each dataset on individual model degrees of freedom. For the higher-order liquid entropy sensitivity shown in Figure 7(a), the peak in the Svechnikov1962 curve was understood to be indicative of a within-dataset initial disagreement in the sign of the log-likelihood gradient with respect to the given parameter. This disagreement was captured by the scaled sensitivity due to the squared gradient term, which increases in magnitude when multiple observations from the same dataset are in conflict. As the apparent conflict was resolved by the MCMC optimization, the sensitivity decreased. Initial sensitivity contributed by some of the other datasets was observed, but quickly dropped to zero as the phase mis-prediction was resolved by the optimization. The log-likelihood contribution of the Bechtoldt1961 dataset remained very sensitive to the regular solution parameter for the A1 phase (Figure 7(b)), and the sensitivity actually increased with respect to the MCMC iterations. While optimization reduces the total log-likelihood gradient, it does not guarantee sensitivity reduction with respect to every dataset.

It is desirable to determine whether an MCMC-based optimization has run for a sufficient number of iterations, and if its estimate of parameter uncertainty is reasonable. In this work the developed Calphad sensitivity theory was applied to perform an analysis using the well-known Cramér–Rao (CR) lower bound on the parameter covariance. The CR bound is a statement on the covariance of an unbiased estimator, giving the inverse of the Fisher information matrix (Eq. **10**) as the lower bound [26]. While any realized estimator may fail to achieve the lower limit, a corollary to the CR bound is that, if a given estimator's covariance falls *below* the given limit, the estimator must be biased in some way.

Corner plots for the A1 and liquid phases, with estimated CR covariance ellipsoids, are shown in Figure 8. For computation of the expectation of the log-likelihood Hessian (Eq. 10), a likelihood-weighted average of the Hessians of the last 300 MCMC iterations, marginalized over all chains, was used. For the model degrees of freedom in the A1 phase, the covariances computed from the MCMC samples were found to be in reasonable agreement with the CR bound, including the reproduction of key correlations. For the liquid model degrees of freedom, MCMC covariances far below the CR bound were observed in the higher-order liquid parameters, indicating bias in the MCMC covariance estimate. This was understood to be caused by an insufficient number of MCMC samples to capture a relatively flat likelihood along those degrees of freedom, in the neighborhood of the maximum-likelihood estimate.

In seeking to remove undesired bias in the liquid uncertainty estimate, two approaches were considered. The first was to use informative priors for the parameters that were found to be insensitive. This would mean adding information from another source, possibly based on experience or intuition. While this could resolve the issue in one sense, the choice of any particular prior would be difficult to justify in advance.

Another approach would be to add more informative observations to the optimization. In this work only phase equilibria measurements were considered, but a pair of liquid mixing enthalpy measurements, for example, would strongly increase both the magnitude of the likelihood gradient (Eq. 8) for the higher-order constant binary interaction term for the liquid, as well as the curvature of the likelihood function along that direction (Eq. 10). The extent of the statistical information for a given model contained within a set of observations can be quantified by the eigenvalues of Eq. 10, and are quantified for two cases in Table 1. One scalar measure based on this spectral approach is the ratio of the largest to smallest eigenvalue (condition number), and was also computed in the table. Assume that such measurements were perfectly consistent with the candidate model (zero error). Even in making the relatively conservative assumption of $\sigma_p = 1000\ J/mol$ for this hypothetical measurement, the strongly informative nature of thermochemical measurements provides a reduction to the uncertainty bound of the higher-order liquid parameters, driving an increase in the smallest FIM eigenvalues and an order-of-magnitude reduction in the matrix condition number. One would then expect subsequent MCMC optimization to be accelerated by the greater curvature of the augmented likelihood function, and an uncertainty estimate closer to the CR bound to be achieved.

It is promising for the future of Calphad sensitivity that this analysis was able to quantifiably reproduce the long-respected wisdom in the Calphad community that thermochemical measurements are the foundation of an accurate thermodynamic model, with the phase diagram playing a highly-visible, yet merely supporting, role.

**Conclusions**

Sensitivity analysis is a powerful tool for the development of Calphad-based thermodynamic models, providing data-point level resolution on the coupling of prediction error to the model parameters. Through the presented theory it was shown possible, in a case study of the Cr-Ni system, to assess the accuracy of MCMC-based covariance estimates using the classical Cramér–Rao bound. Computation of the Fisher information matrix quantified the statistical value of thermochemical measurements versus phase equilibria measurements; the former was shown to be much greater. Further analysis of the role of conflicting data in multi-component Calphad model sensitivity, and how it might influence the design of new experiments, was suggested for future work.

**Acknowledgments**

This research was carried out at the Jet Propulsion Laboratory, California Institute of Technology, under a contract with the National Aeronautics and Space Administration. BB and ZKL were supported by a NASA Space Technology Research Fellowship grant number 80NSSC18K1168.

# References

[1] M. Stan, B.J. Reardon, A Bayesian approach to evaluating the uncertainty of thermodynamic data and phase diagrams, Calphad. 27 (2003) 319–323. https://doi.org/10.1016/j.calphad.2003.11.002.

[2] R.A. Otis, Z.-K. Liu, High-Throughput Thermodynamic Modeling and Uncertainty Quantification for ICME, JOM. 69 (2017) 886–892. https://doi.org/10.1007/s11837-017-2318-6.

[3] T.C. Duong, R.E. Hackenberg, A. Landa, P. Honarmandi, A. Talapatra, H.M. Volz, A. Llobet, A.I. Smith, G. King, S. Bajaj, A. Ruban, L. Vitos, P.E.A. Turchi, R. Arróyave, Revisiting thermodynamics and kinetic diffusivities of uranium–niobium with Bayesian uncertainty analysis, Calphad. 55 (2016) 219–230. https://doi.org/10.1016/j.calphad.2016.09.006.

[4] P. Honarmandi, R. Arróyave, Uncertainty Quantification and Propagation in Computational Materials Science and Simulation-Assisted Materials Design, Integr Mater Manuf Innov. 9 (2020) 103–143. https://doi.org/10.1007/s40192-020-00168-2.

[5] N.H. Paulson, B.J. Bocklund, R.A. Otis, Z.-K. Liu, M. Stan, Quantified uncertainty in thermodynamic modeling for materials design, Acta Materialia. 174 (2019) 9–15. https://doi.org/10.1016/j.actamat.2019.05.017.

[6] B. Jansson, Evaluation of Parameters in Thermochemical Models Using Different Types of Experimental Data Simultaneously, Royal Institute of Technology, Stockholm, 1984.

[7] B. Sundman, X.-G. Lu, H. Ohtani, The implementation of an algorithm to calculate thermodynamic equilibria for multi-component systems with non-ideal phases in a free software, Computational Materials Science. 101 (2015) 127–137. https://doi.org/10.1016/j.commatsci.2015.01.029.

[8] B. Bocklund, R. Otis, A. Egorov, A. Obaied, I. Roslyakova, Z.-K. Liu, ESPEI for efficient thermodynamic database development, modification, and uncertainty quantification: application to Cu–Mg, MRS Communications. 9 (2019) 618–627. https://doi.org/10.1557/mrc.2019.59.

[9] W. Cao, S.-L. Chen, F. Zhang, K. Wu, Y. Yang, Y.A. Chang, R. Schmid-Fetzer, W.A. Oates, PANDAT software with PanEngine, PanOptimizer and PanPrecipitation for multi-component phase diagram calculation and materials property simulation, Calphad. 33 (2009) 328–342. https://doi.org/10.1016/j.calphad.2008.08.004.

[10] H. Yue, M. Brown, F. He, J. Jia, D.B. Kell, Sensitivity analysis and robust experimental design of a signal transduction pathway system, Int. J. Chem. Kinet. 40 (2008) 730–741. https://doi.org/10.1002/kin.20369.

[11] N.H. Paulson, S. Zomorodpoosh, I. Roslyakova, M. Stan, Comparison of statistically-based methods for automated weighting of experimental data in CALPHAD-type assessment, Calphad. 68 (2020) 101728. https://doi.org/10.1016/j.calphad.2019.101728.

[12] A.T. Dinsdale, SGTE Data For Pure Elements, (1989).

[13] F. Tang, B. Hallstedt, Using the PARROT module of Thermo-Calc with the Cr–Ni system as example, Calphad. 55 (2016) 260–269. https://doi.org/10.1016/j.calphad.2016.10.003.

[14] L.A. Pugliese, G.R. Fitterer, Activities and phase boundaries in the Cr-Ni system using a solid electrolyte technique, Metallurgical Transactions. 1 (1970) 1997–2002. https://doi.org/10.1007/BF02642800.

[15] W. a. Dench, Adiabatic high-temperature calorimeter for the measurement of heats of alloying, Transactions of the Faraday Society. 59 (1963) 1279. https://doi.org/10.1039/tf9635901279.

[16] V.N. Svechnikov, V.M. Pan, Characteristics of the Equilibrium Diagram and Processes of Solution and Precipitation in the Cr- Ni System, Sbornik Naučnych Trudov Instituta Metallofiziki. 15 (1962) 164–178.

[17] M.J. Collins, Electron optic determination of solid phase boundaries in Ni–Cr system, Materials Science and Technology. 4 (1988) 560–561. https://doi.org/10.1179/mst.1988.4.6.560.

[18] A. Watson, F.H. Hayes, Enthalpies of formation of solid NiCr and NiV alloys by direct reaction calorimetry, Journal of Alloys and Compounds. 220 (1995) 94–100. https://doi.org/10.1016/0925-8388(94)06008-8.

[19] P. Saltykov, V.T. Witusiewicz, I. Arpshofen, H.J. Seifert, F. Aldinger, Enthalpy of mixing of liquid Al-Cr and Cr-Ni alloys, Journal of Materials Science and Technology. 18 (2002) 167–170.

[20] Q. Zhang, J.C. Zhao, Impurity and interdiffusion coefficients of the Cr-X (X = Co, Fe, Mo, Nb, Ni, Pd, Pt, Ta) binary systems, Journal of Alloys and Compounds. 604 (2014) 142–150. https://doi.org/10.1016/j.jallcom.2014.03.092.

[21] L. Karmazin, Lattice parameter studies of structure changes of NiCr alloys in the region of Ni2Cr, Materials Science & Engineering, A: Structural Materials: Properties, Microstructure and Processing. 54 (1982) 247–256. https://doi.org/10.1016/0025-5416(82)90119-7.

[22] U. Thiedemann, M. Rösner-Kuhn, D.M. Matson, G. Kuppermann, K. Drewes, M.C. Flemings, M.G. Frohberg, Mixing enthalpy measurements in the liquid ternary system iron-nickel-chromium and its binaries, Steel Research. 69 (1998) 3–7. https://doi.org/10.1002/srin.199801599.

[23] C.J. Bechtoldt, H.C. Vacher, Redetermination of the chromium and nickel solvuses in the chromium-nickel system, TRANSACTIONS OF THE METALLURGICAL SOCIETY OF AIME. 221 (1961) 14–18.

[24] C.H.M. Jenkins, E.H. Bucknall, C.R. Austin, G.A. Mellor, Some alloys for use at high temperatures: Part IV: The constitution of the alloys of nickel, chromium and iron, Journal of the Iron and Steel Institute. 136 (1937).

[25] A. Taylor, R.W. Floyd, The constitution of Nickel-rich alloys of the Nickel-Chromium-Titanium system, Journal of the Institute of Metals. 80 (1952) 577–587.

[26] C.R. Rao, Information and the Accuracy Attainable in the Estimation of Statistical Parameters, in: S. Kotz, N.L. Johnson (Eds.), Breakthroughs in Statistics: Foundations and Basic Theory, Springer, New York, NY, 1992: pp. 235–247. https://doi.org/10.1007/978-1-4612-0919-5_16.

**Figures**

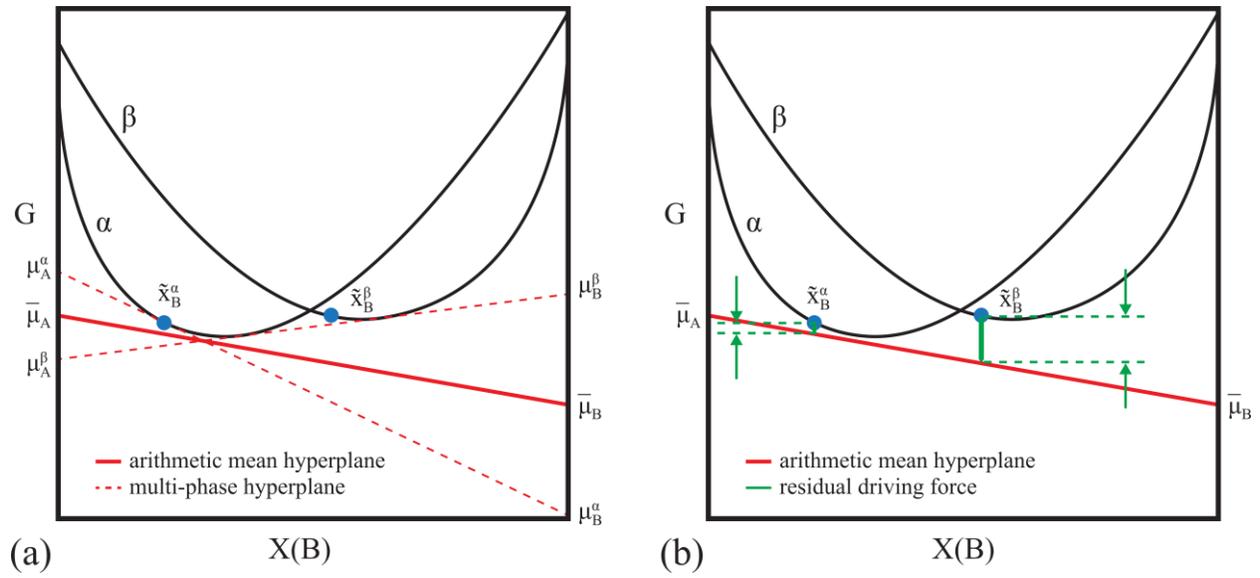

**Figure 1. Residual driving force $R(\theta)$ of a phase co-existence measurement (Eq. 1).**

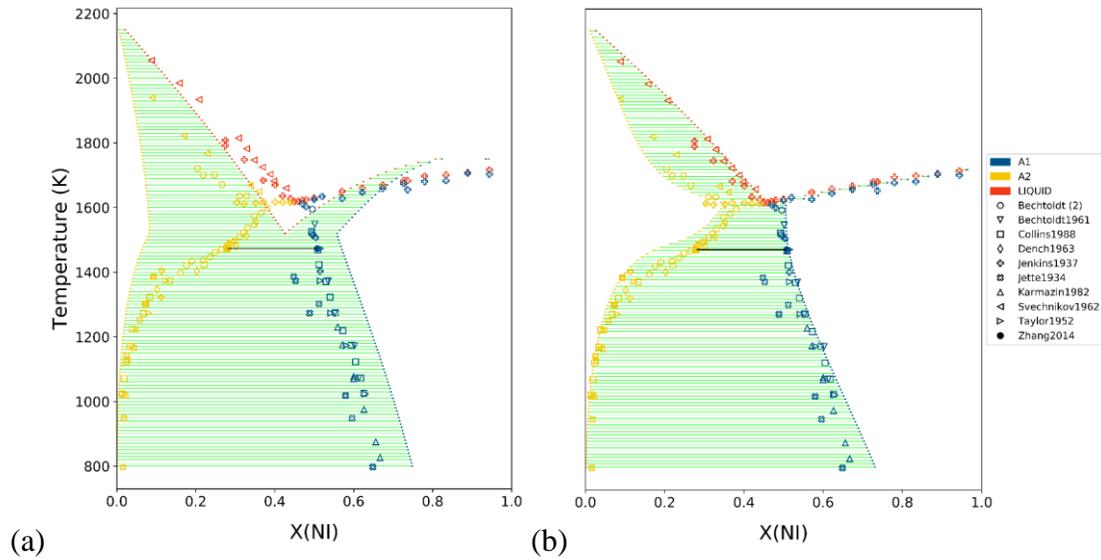

**Figure 2. Initial and final phase diagram of Cr-Ni, with experimentally measured phase equilibria from the literature superimposed.** (a) The initial Cr-Ni thermodynamic model was generated by the ESPEI software using thermochemical data (not shown) for the individual phases. (b) The Cr-Ni phase diagram is shown after 500 Markov Chain Monte Carlo iterations, using only the shown phase equilibria measurements as input.

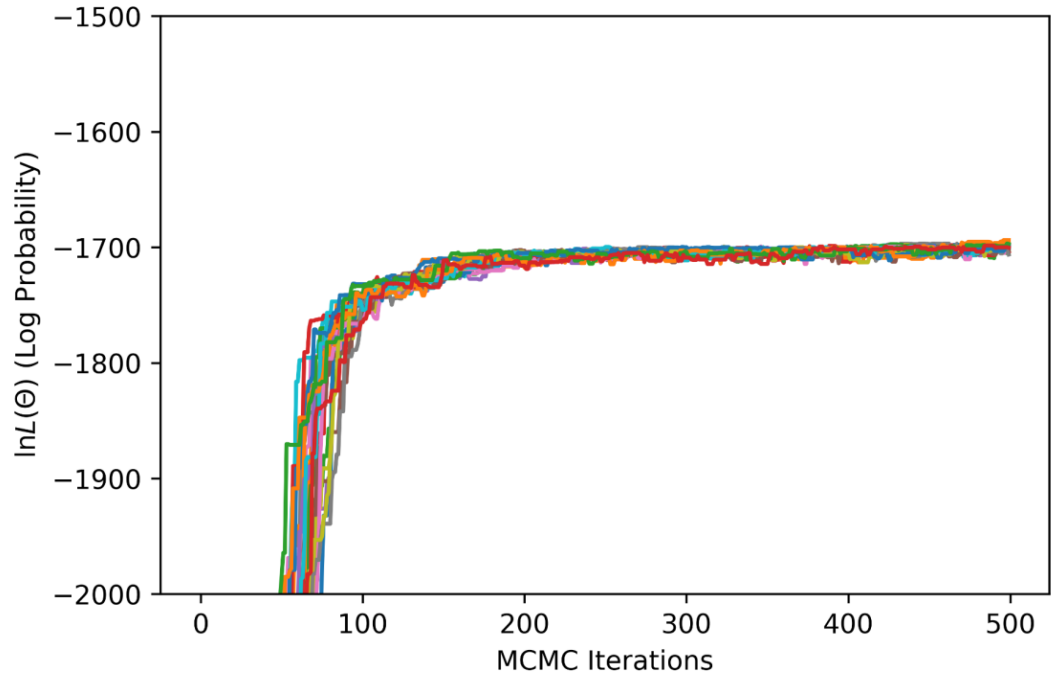

**Figure 3. ESPEI Markov Chain Monte Carlo (MCMC) log-likelihood trace**.

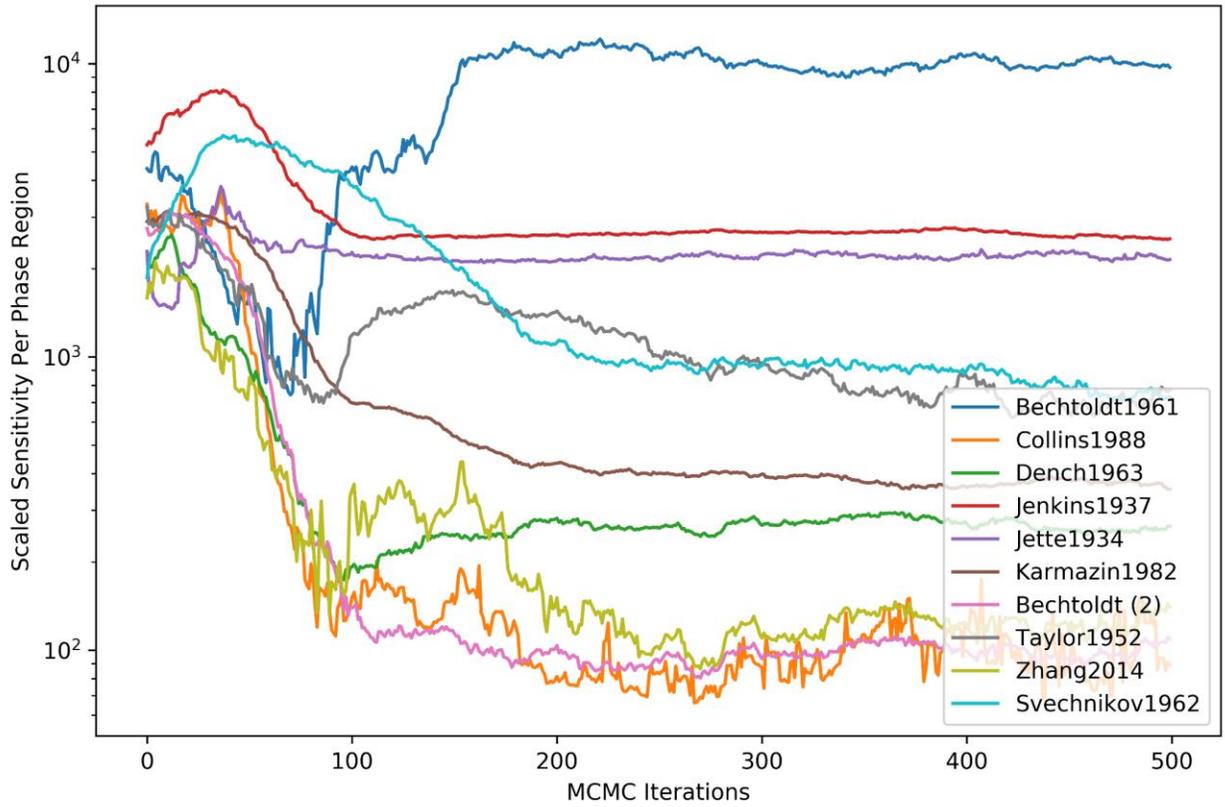

**Figure 4**. **Dataset scaled sensitivity per phase region.** The scaled sensitivity (Eq. 12) was computed as a summation over all parameters ($m$) and observations ($p$) contained within each dataset, normalized based on the number of contained measurements (phase regions).

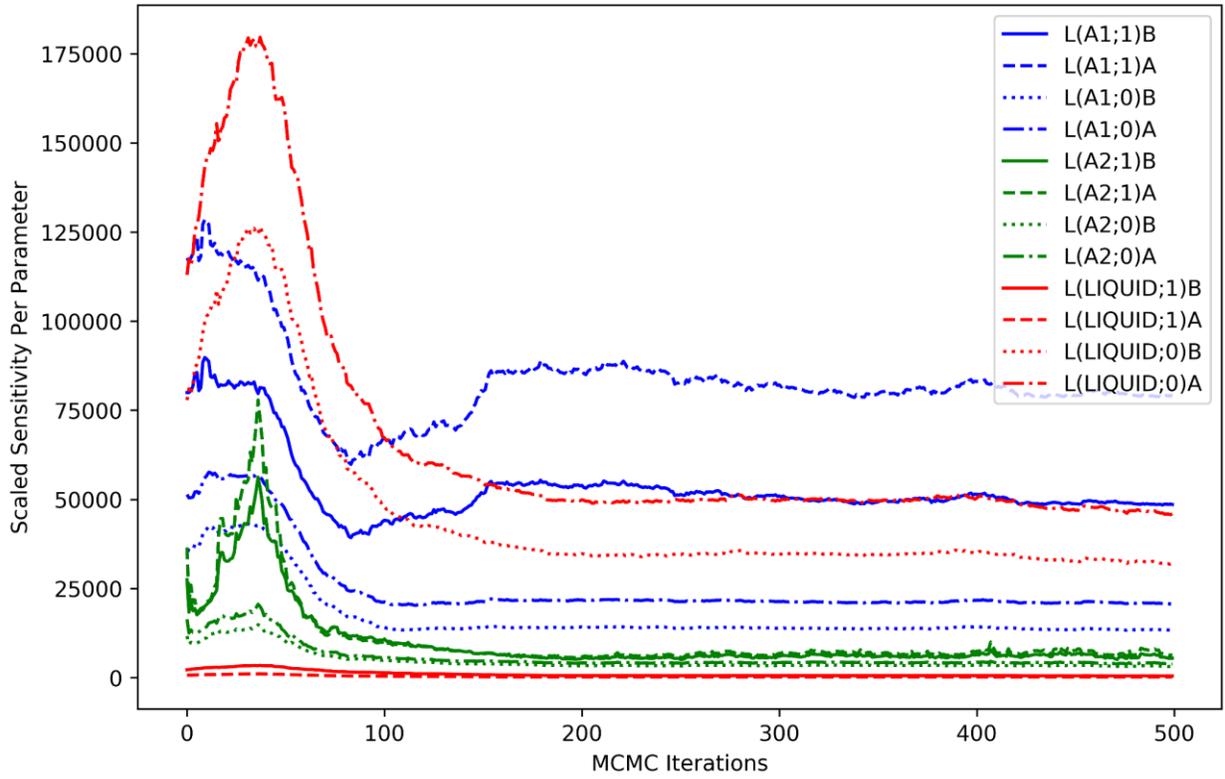

**Figure 5**. **Scaled sensitivity per parameter.** The contribution of each parameter ($m$) to the scaled sensitivity (Eq. **12**) is computed as a summation over all observations ($p$) in all datasets, and is shown as a function of Markov Chain Monte Carlo (MCMC) iterations.

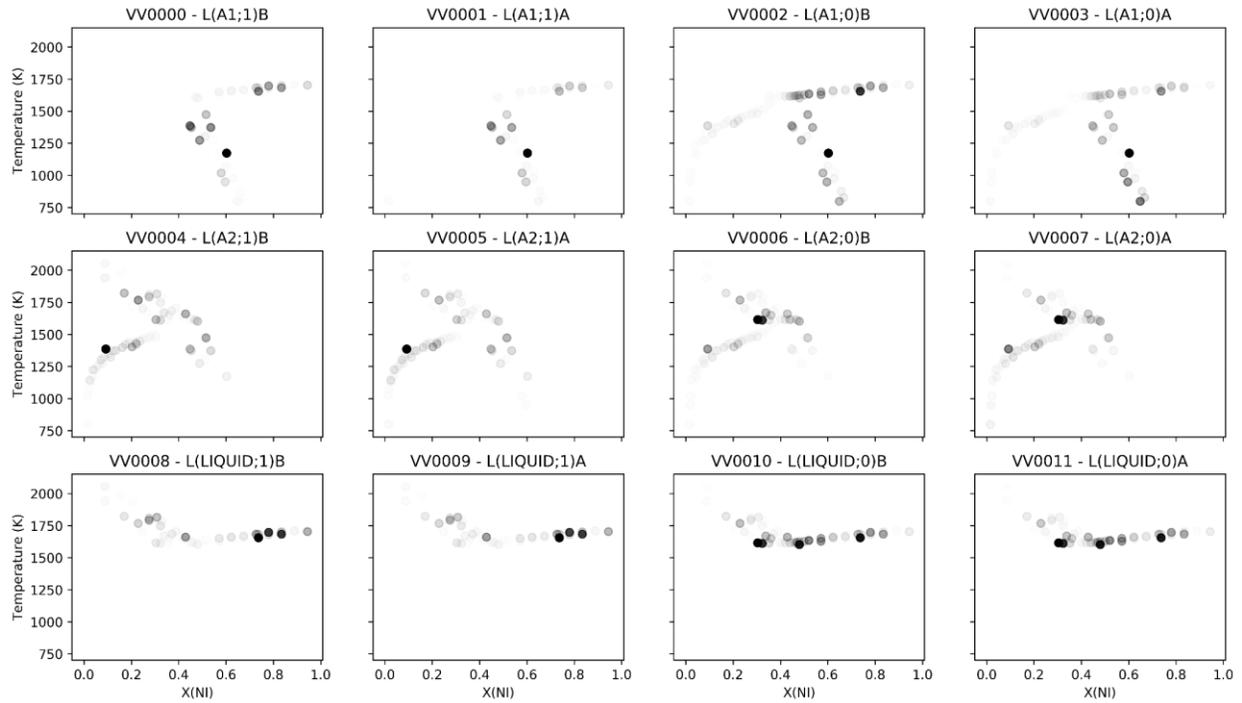

**Figure 6. Scaled sensitivity per parameter averaged over the last 300 MCMC iterations, visualized in the space of observations.** Each subplot was separately normalized, such that full opacity corresponded to the largest observed scaled sensitivity for the given parameter.

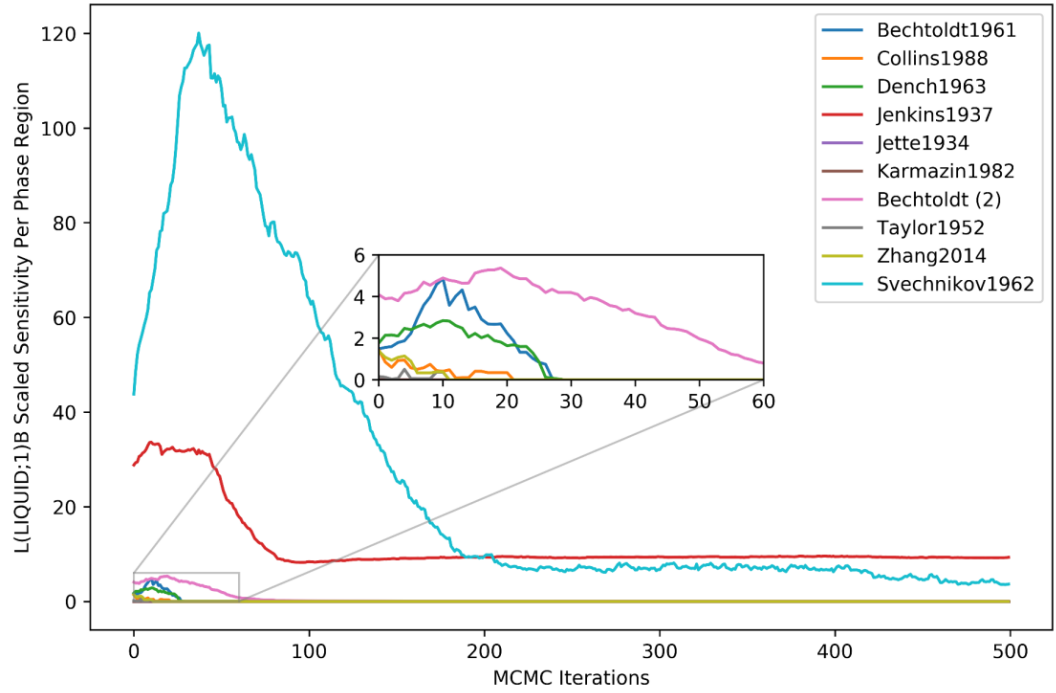

(a)

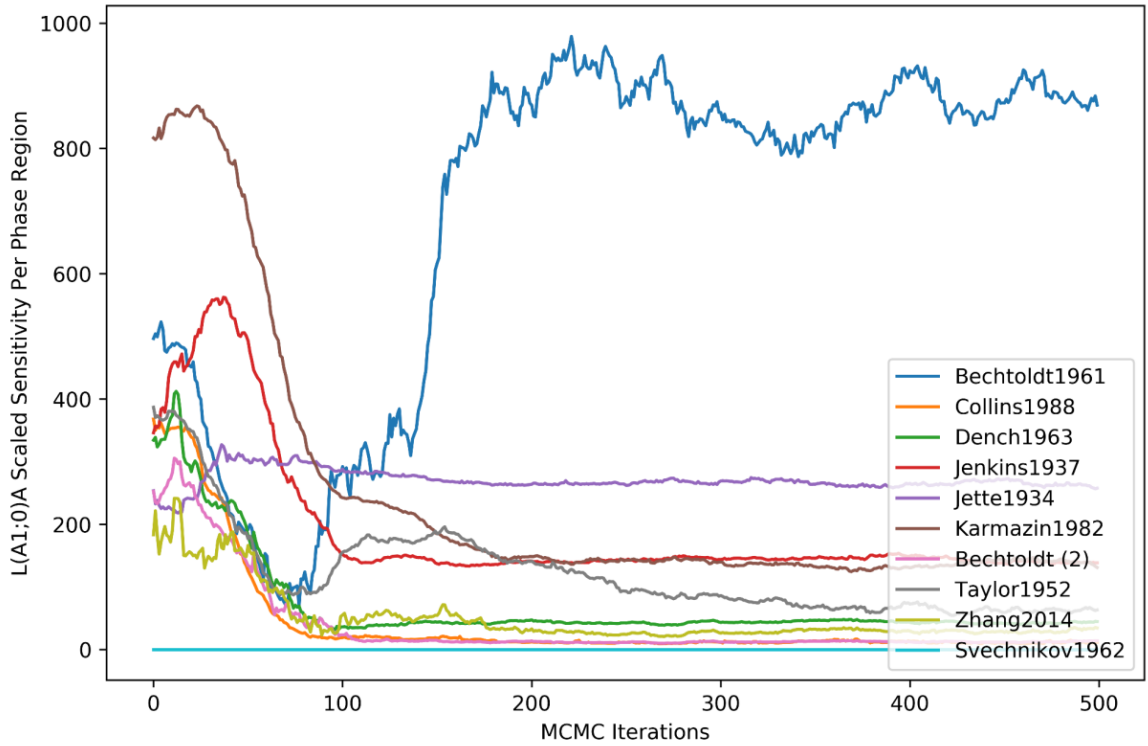

(b)

**Figure 7. Parameter scaled sensitivity per dataset.** (a) Higher-order entropy parameter of the liquid and (b) regular solution parameter of the A1 phase.

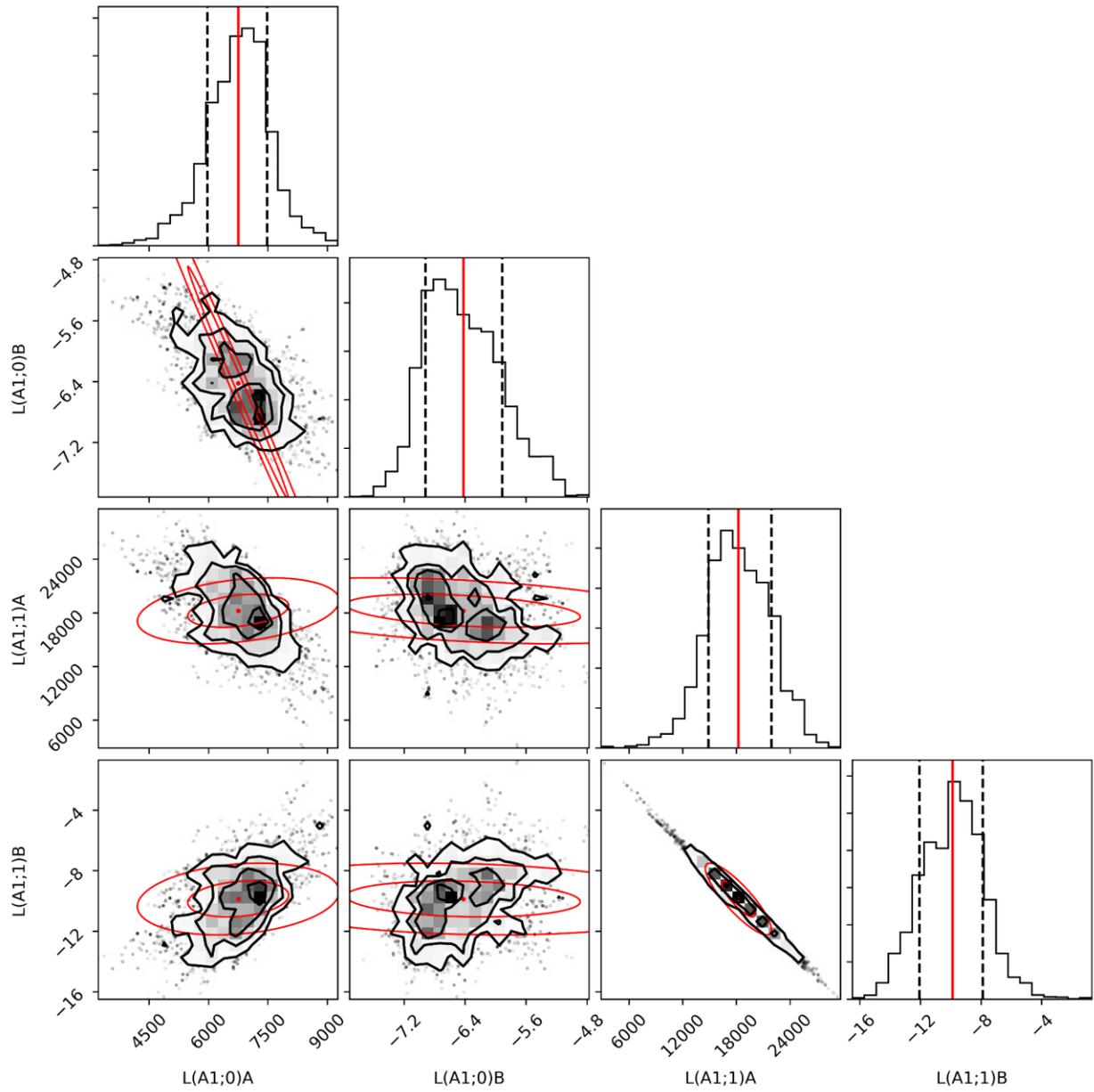

(a)

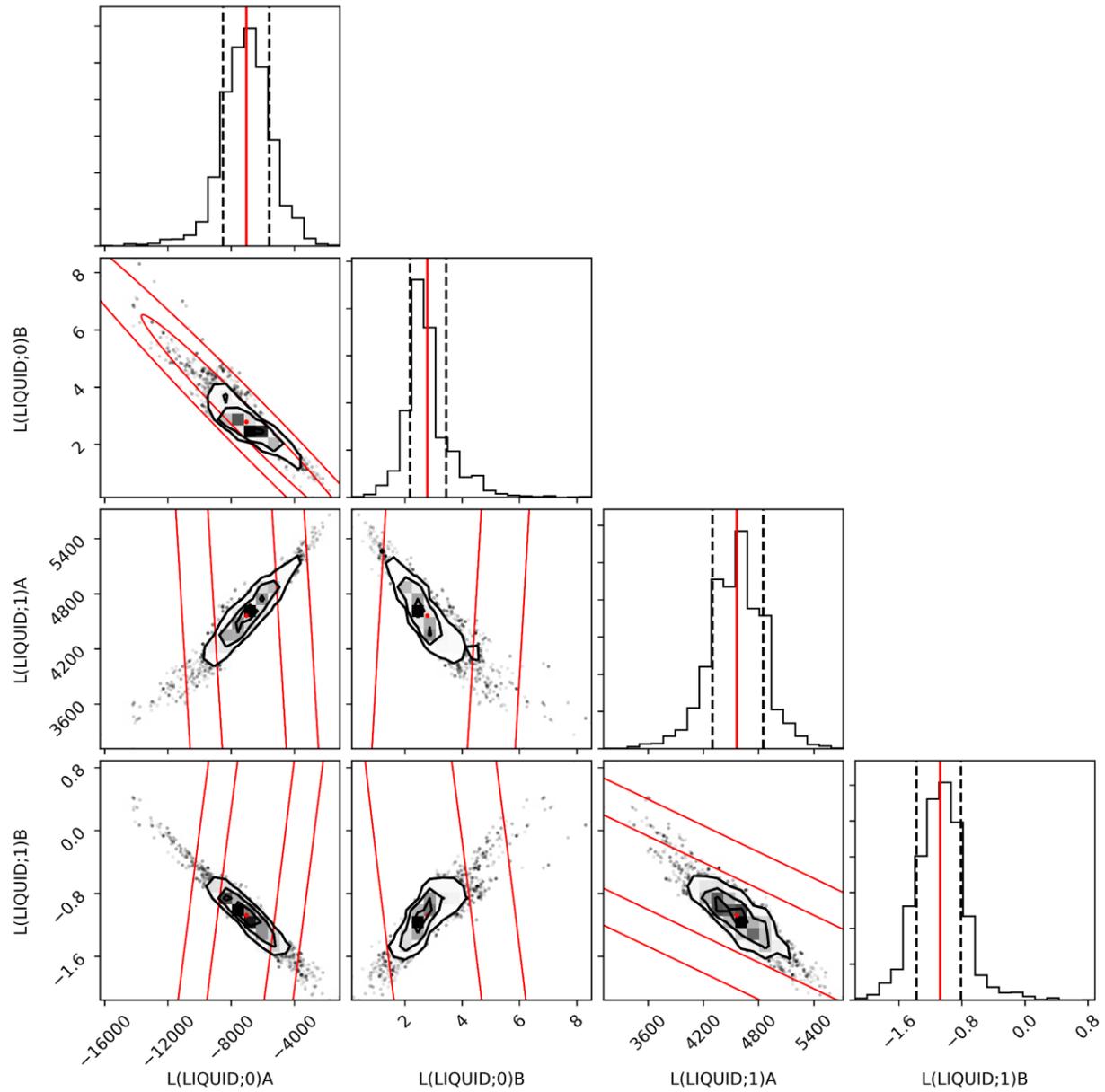

(b)

**Figure 8. Corner plots for (a) the A1 and (b) liquid phase, with estimated Cramér–Rao (CR) covariance ellipsoids superimposed in red, at 1 and 2 standard deviations.**

**Table 1.** Eigenvalues of the estimated Fisher information matrix (FIM), before and after addition of two hypothetical liquid enthalpy measurements to the likelihood function.

|  | $\lambda_{max}$ | $\lambda_{min}$ | $\lambda_{max}/\lambda_{min}$ |
|---|---|---|---|
| **Phase Equilibria Only** | $1.86 \times 10^5$ | $2.70 \times 10^{-9}$ | $6.87 \times 10^{13}$ |
| **Including Enthalpy Measurements** | $1.86 \times 10^5$ | $2.68 \times 10^{-8}$ | $6.93 \times 10^{12}$ |

The full Supplementary Material for this manuscript, including all code needed to reproduce the figures and table, was too large to directly include here. It can be found on the Open Science Framework digital repository at the following link.

https://osf.io/mxqfs/?view_only=a9a9c23a89554aee8f7a8e4914ca5e0b

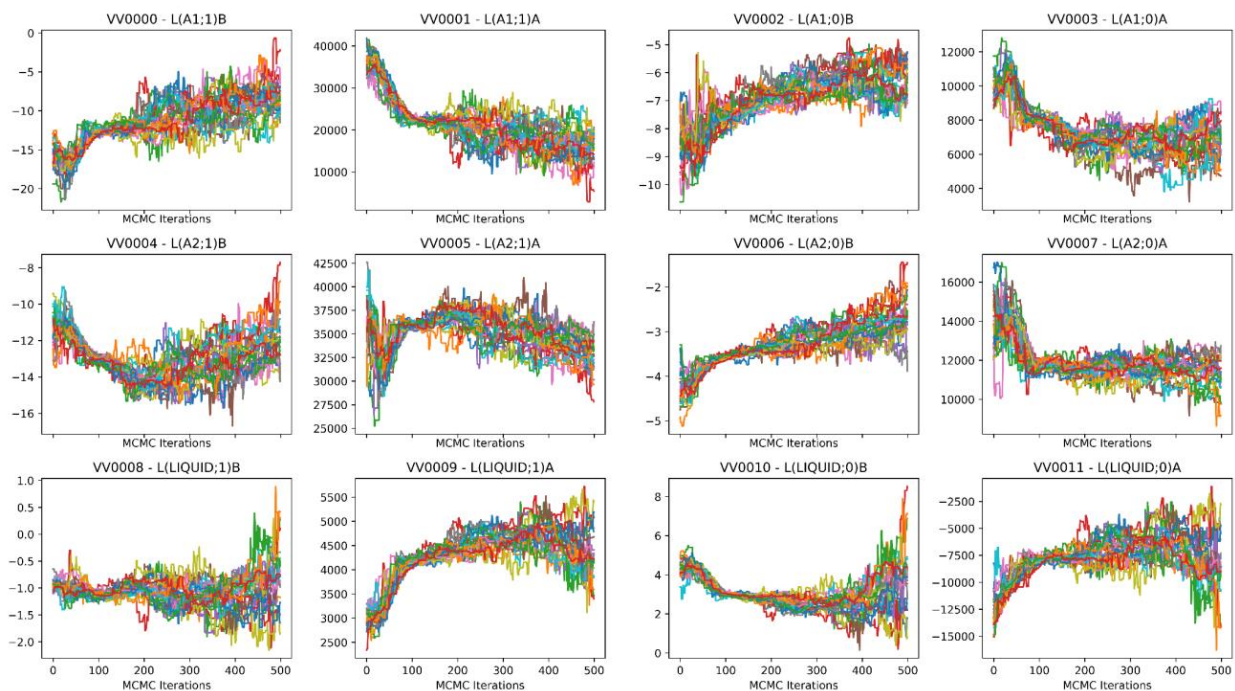

**Supplementary Figure. ESPEI Markov Chain Monte Carlo (MCMC) parameter trace.** Reasonable agreement between MCMC chains in the ensemble was observed, though there was some indication that the stationary distribution may not yet have been achieved for some of the liquid parameters, despite identification of a maximum-likelihood configuration which produced a satisfactory phase diagram.